\documentclass[12pt]{amsart}
\usepackage[cp1251]{inputenc}
\usepackage[english]{babel}
\usepackage{amsmath}
\usepackage{amssymb}
\usepackage{amsfonts}
\usepackage{epsfig}
\usepackage{srcltx}
\usepackage{cite}
\usepackage{float}
\usepackage[ a4paper,  mag=1000, includefoot,  left=2cm, right=2cm, top=2cm, bottom=2cm, headsep=1cm, footskip=1cm ]{geometry}
\usepackage[unicode,colorlinks]{hyperref}
\hypersetup{ colorlinks=true, citecolor=blue, linkcolor=blue}

\newtheorem{Th}{Theorem}
\newtheorem{Def}{Definition}

\begin{document}

\thispagestyle{empty}

\title{Stability of parametric autoresonance under random perturbations}

\author{O. Sultanov}

\address{Oskar Sultanov,
\newline\hphantom{iii}  Institute of Mathematics USC RAS
\newline\hphantom{iii}  112 Chernyshevsky str., Ufa 450008, Russia}

\email{oasultanov@matem.anrb.ru}

\maketitle {\small
\begin{quote}


\end{quote}
\begin{quote}
\noindent{\bf Abstract.}
A mathematical model describing the initial stage of the capture into the parametric autoresonance in nonlinear oscillating systems with a dissipation is considered. Solutions with unboundedly growing energy in time at infinity are associated with the autoresonance phenomenon. Stability of such solutions is investigated. We describe classes of admissible deterministic and random perturbations such that the stability of autoresonance is preserved on an asymptotically large interval.
\medskip

\end{quote} }

\section{Introduction}

Autoresonance is a phenomenon of a considerable growth of the energy of forced nonlinear systems. This phenomenon plays an important role in a wide class of various physical problems associated with nonlinear oscillations and waves~\cite{Veksl44,McMillan45,Fridl01}. The majority of well-known theoretical investigations \cite{Khain01,LK08,OKSG07,Fr09,GKT10,Fridl14} consists a numerical and asymptotic analysis of mathematical models describing the initial stage of a capture into autoresonance. However, the stability of such models in the presence of external perturbations remained an open question. In the present work we study the problem of stability of parametric autoresonance in nonlinear systems under persistent perturbations.

Consider a model system of primary parametric resonance equations~\cite{Khain01}:
\begin{equation}
    \begin{array}{c}
        {\displaystyle
            \frac{dr}{d\tau}= r\sin\psi-\delta r,
        }
        \quad
        {\displaystyle
            \frac{d\psi}{d\tau}=r-\lambda \tau + f \cos\psi, \quad \tau>0.
        }
    \end{array}
    \label{Main_Res}
\end{equation}
This system appears in the asymptotic analysis of nonlinear oscillations driven by a small force.
The unknown functions $r(\tau)$ and $\psi(\tau)$ represent the slow varying amplitude (energy) and phase shift of fast harmonic oscillations. The parameters $\lambda>0$ and $f\neq 0$ are factors related to the driving frequency and amplitude. The positive constant $\delta$ correspond to dissipation coefficient. Solutions with unboundedly growing energy in time are associated with the capture of an oscillatory nonlinear system into parametric autoresonance. The aim of this paper to prove the stability of resonance solutions. It is assumed that only stable solutions correspond to motions that are observed in nature. Note that stability of the autoresonance in nondissipative systems ($\delta=0$) was discussed in~\cite{OS14}.

\begin{figure}[H]\label{F1}
\centering
\includegraphics[width=0.45\textwidth]{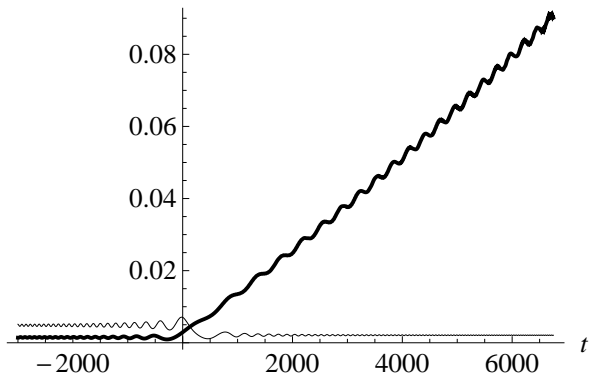}
\caption{ The energy of the parametrically driven nonlinear oscillator \eqref{Duffing}. Parameter values are $\beta=0$, $\gamma=3/2$, $\varepsilon = 0.001$, $\alpha = \varepsilon^2/8$. Two curves correspond to different initial points.}
\end{figure}

System \eqref{Main_Res} is derived by an averaging of parametrically driven nonlinear oscillations~\cite{BogMitr61}. Let us consider the equation
\begin{equation}\label{Duffing}
    {\displaystyle
        \frac{d^2x}{dt^2}+ \beta \frac{dx}{dt}+(1+\varepsilon \cos\phi)\,x+\gamma x^3=0
     }
\end{equation}
as an example of the initial mathematical model. Here, $\phi(t;\alpha)=2t+\alpha t^2$, $0<\varepsilon,\alpha,\beta\ll 1$, $\gamma=$ const $>0$. The point $x=0$ is a stable equilibrium of the unperturbed oscillator ($\varepsilon=0$). Solutions of equation \eqref{Duffing} with initial values near the equilibrium $|x(0)|+|\dot x(0)|\ll 1$ and with amplitudes increasing up to the order of unity at large times correspond to autoresonance (see Figure~\ref{F1}).

\noindent
The asymptotic of resonance solutions of equation \eqref{Duffing} for $0\leq t\ll \varepsilon^{-2}$ is constructed in the form:
$$
    x(t)=\sqrt{\kappa \varepsilon r (\tau) } \cos\frac12 (\phi+\psi(\tau))+\mathcal{O}(\varepsilon), \quad \varepsilon\to 0, \quad \tau=\frac{\varepsilon t}{2}, \quad \kappa=\frac{2}{3\gamma},
$$
where the slow varying functions $r(\tau)$ and $\psi(\tau)$ satisfy system \eqref{Main_Res} with $\lambda=8\alpha \varepsilon^{-2}$, $\delta=2\beta \varepsilon^{-1}$, and $f=1$.

\section{Autoresonance solutions}
The solutions of system \eqref{Main_Res} cannot be written in an explicit form. However, the asymptotic solutions with increased energy at infinity $\tau\to\infty$ can be constructed in the form of power asymptotic series with constant coefficients:
\begin{equation}\label{Asympt}
    {\displaystyle
    R_\pm(\tau)=\lambda \tau + \sum_{j=0}^{\infty}  r^\pm_j\, \tau^{-j}, \quad
    \Psi_\pm(\tau)=\sum_{j=0}^{\infty}\psi^\pm_j\, \tau^{-j}, \quad \tau\to\infty.
    }
\end{equation}
Substituting these series in system \eqref{Main_Res} and equating the expressions of the same powers give the
recurrence relations for determining the coefficients $r^\pm_j$ and $\psi^\pm_j$.
In this way two solutions are constructed if $0<\delta<1$; difference stem from two roots of the trigonometric equation $\sin\psi^\pm_0=\delta$:
$$
    \psi^\pm_1=\frac{1}{\cos\psi^\pm_0}, \quad r^\pm_0=-f\cos\psi^\pm_0, \quad r^\pm_1=f \tan \psi^\pm_0.
$$
Series \eqref{Asympt} correspond to the exact solutions of system \eqref{Main_Res} with a given asymptotic expansion at infinity~\cite{Kuzn89}. We investigate the stability of the solutions $R_\pm(\tau)$, $\Psi_\pm(\tau)$ in the sense of Lyapunov and under persistent perturbations.

The solution with the phase $\psi^+_0=\arcsin \delta$ is unstable as can be seen by analyzing the equations linearized near the leading term of the asymptotic solution: one of the characteristic roots has a positive real part. In the case of $\psi^-_0=\pi-\arcsin\delta$, such an approach is unapplicable because the characteristic roots are purely imaginary. In this situation the property of stability depends on non-linear and time-dependent terms of equations (see~\cite{LKOS13,OS10}).
\medskip

\section{Perturbed equations}
Along with \eqref{Main_Res}, we consider the perturbed system
\begin{equation}\label{PSD}
    \begin{array}{c}
        {\displaystyle \frac{dr}{d \tau}=(1+\mu \xi)r\sin\psi-\delta r, \ \  \frac{d\psi}{d \tau}=r-\lambda \tau+\mu \zeta+(f+ \mu \eta)\cos\psi,}
    \end{array}
\end{equation}
where the perturbations $\xi(r,\psi,\tau)$, $\eta(r,\psi,\tau)$, and $\zeta(r,\psi,\tau)$ are defined for $(r,\psi)\in \mathbb R^2$, $\tau>0$. The coefficient $\mu\in\mathbb R$, $0<\mu\ll 1$ is a perturbation parameter. We consider the perturbations such that system \eqref{PSD} has a global solution. To ensure this property we have to require additional restrictions (see~\cite{NemStep04,Hasm69}) on the class of functions $(\xi,\eta,\zeta)$. Our goal is to identify a class of persistent perturbations $\mathcal P$ such that any solution $r_\mu(\tau)$, $\psi_\mu(\tau)$ of system \eqref{PSD} with initial data from a neighborhood of \eqref{As} remains near the solution $R_-(\tau)$, $\Psi_-(\tau)$ while the parameter $\mu$ is small and $(\xi,\eta,\zeta)$ belong to $\mathcal P$.

The example of original system leading to \eqref{PSD} is the Duffing oscillator
\begin{equation}\label{PDuf}
    {\displaystyle
        \frac{d^2x}{dt^2}+\beta\frac{dx}{dt}+\Big\{1+\varepsilon(1+\mu a)\cos(\phi+\mu \varphi)\Big\}\,x+\gamma x^3=0, \quad 0<\mu\ll1,
     }
\end{equation}
where the functions $a(x,\dot x,t;\varepsilon)$ and $\varphi(x,\dot x,t;\varepsilon)$ correspond to perturbations of the pumping amplitude and phase. If $a=a(t;\varepsilon)$ and $\varphi=\varphi(t;\varepsilon)$, then the perturbations in system \eqref{PSD} have the form
\begin{equation}\label{PF}
    {\displaystyle\xi(\tau)=a(t;\varepsilon), \ \ \eta(\tau)=a(t;\varepsilon), \ \ \zeta(\tau)=-\frac{4 \varphi'_t(t;\varepsilon)}{ \varepsilon}, \ \ t=\frac{2 \tau}{\varepsilon}.}
\end{equation}

\section{Lyapunov stability}
To study the stability of the solution $R_-(\tau)$, $\Psi_-(\tau)$, we use the first terms of asymptotic expansion \eqref{Asympt}
\begin{equation}\label{As}
    {\displaystyle
R_-(\tau)=\lambda \tau+f \sigma+\mathcal O(\tau^{-1}), \quad \Psi_-(\tau)=\pi-\arcsin\delta-\frac{1}{\sigma}\tau^{-1}+\mathcal O(\tau^{-2}),
}
\end{equation}
where $\sigma=\sqrt{1-\delta^2}$.
\begin{Th}
\label{Th1}
If\, $0<\delta<1$ and $f>0$, then the solution $R_-(\tau)$, $\Psi_-(\tau)$ with asymptotics \eqref{As} is asymptotically stable.
\end{Th}
\noindent
{\bf Proof.} By the change of variables
\begin{equation}\label{Exch1}
    {\displaystyle
    r=R_-(\tau)+\sqrt{\lambda \tau}\, R, \quad
    \psi=\Psi_-(\tau)+\Psi
    }
\end{equation}
system \eqref{Main_Res} can be rewritten in the form
\begin{equation}\label{AlmHam}
\begin{array}{c}
    {\displaystyle
        \frac{1}{\sqrt{\lambda\tau}}\frac{d R}{d\tau}=-\partial_\Psi H(R,\Psi,\tau)+F(R,\Psi,\tau),
    }\quad
    {\displaystyle
       \frac{1}{\sqrt{\lambda\tau}} \frac{d\Psi}{d\tau}=\partial_R H(R,\Psi,\tau),
    }
\end{array}
\end{equation}
where
\begin{align}
       & {\displaystyle H (R,\Psi,\tau) = \frac{R^2}{2}+ \frac{R_-}{\lambda \tau} \Big[\cos(\Psi+\Psi_-)-\cos\Psi_-+\Psi\sin\Psi_-\Big] + \frac{f R}{\sqrt{\lambda \tau}}\Big[\cos(\Psi+\Psi_-)-\cos\Psi_-\Big], }
       \nonumber
       \\
       & {\displaystyle  F (R,\Psi,\tau) =-\frac{R}{\sqrt{\lambda\tau}} \Big[\delta+(f-1)\sin(\Psi+\Psi_-)\Big]-\frac{R}{2}\tau^{-1}. }
    \nonumber
\end{align}
For the new functions $R(\tau)$, $\Psi(\tau)$ we study the problem of stability of the equilibrium $(0;0)$  by the Lyapunov second method. To construct a Lyapunov function for system \eqref{AlmHam} the asymptotics of the right-hand sides in a neighborhood of the equilibrium (as $\rho=\sqrt{R^2+\Psi^2}\to 0$) and at infinity (as $\tau\to\infty$) are used. Note that all asymptotic estimates written out bellow in the form $\mathcal{O}(\rho^n)$ and $\mathcal{O}(\tau^{-m})$ ($n,m={\hbox{\rm const}}>0$) are uniform with respect to $R,\Psi,\tau$ in the domain
$$
    \mathcal B(\rho_\ast,\tau_\ast)=\{(R,\Psi,\tau): \rho<\rho_\ast, \, \tau>\tau_\ast\}, \quad \rho_\ast,\tau_\ast={\hbox{\rm const}}>0.
$$

It can easily be checked that the Hamiltonian has a positive quadratic form as the leading term of the asymptotic expansion:
$$
    H=\frac{R^2}{2}+\sigma\frac{\Psi^2}{2}+\mathcal{O}(\rho^3)+\mathcal{O}(\rho^2)\mathcal{O}(\tau^{-1/2}), \quad \rho\to0, \quad \tau\to\infty.
$$
By taking into account \eqref{As} one can readily write out asymptotics of the derivatives:
$$\begin{array}{lcl}
   {\displaystyle \partial_R H }& = &{\displaystyle R+  \frac{f}{\sqrt{\lambda\tau}} \big[\sigma (1-\cos\Psi)-\delta\sin\Psi \big]
+\mathcal{O}(\rho)\mathcal{O}(\tau^{-3/2}),  }
    \\ \\
 {\displaystyle
    \partial_\Psi H }&=& {\displaystyle \sigma\,\sin\Psi+\delta (1-\cos\Psi)+}  {\displaystyle  \frac{R}{\sqrt{\lambda\tau}} \big[ \sigma \sin\Psi+\delta (1-\cos\Psi)\big]+\mathcal{O}(\rho)\mathcal{O}(\tau^{-1}),}
 \\ \\
    {\displaystyle \partial_\tau H }& = &{\displaystyle \mathcal{O}(\rho^2)\mathcal{O}(\tau^{-3/2}).}
\end{array}$$
The non-Hamiltonian part $F(R,\Psi,\tau)$ tends to zero as $\tau\to\infty$:
$$
    F=-m \big[R+\mathcal O(\rho^2)\big] \tau^{-1/2}+\mathcal O(\rho)\mathcal O(\tau^{-1}), \quad m=\frac{\delta f}{\sqrt\lambda}>0.
$$

The Lyapunov function is constructed on the basis of the Hamiltonian:
\begin{equation}\label{LF}
    {\displaystyle
    V(R,\Psi,\tau)=H(R,\Psi,\tau)+ \frac{m}{2}  R \Psi \, \tau^{-1/2}.}
\end{equation}
The derivative of the function $V(R,\Psi,\tau)$ along the trajectories of system \eqref{AlmHam} decreases as $\tau\to\infty$; the leading term of its asymptotic expansion consists a quadratic form:
$$\begin{array}{rll}
        {\displaystyle \frac{1}{\sqrt{\lambda\tau}}\frac{dV}{d\tau}\Big|_{\eqref{AlmHam}} }
        & = &
        {\displaystyle \frac{1}{\sqrt{\lambda\tau}}\frac{\partial V}{\partial \tau}+\frac{\partial V}{\partial R}\Big[-\frac{\partial H}{\partial \Psi}+ F\Big] +\frac{\partial V}{\partial \Psi}\frac{\partial H}{\partial R}=    }
        \\ \\
        {\displaystyle }
        & = &
        {\displaystyle   -\frac{m}{2} \big[R^2+\sigma\,\Psi^2\big] [1+\mathcal O(\rho)+\mathcal O(\tau^{-1})] \tau^{-1/2}. }
\end{array}$$
Note that the remainders can be made arbitrarily small by choosing suitable domain $\mathcal B(\rho_\ast,\tau_\ast)$. It follows that there exist $\rho_1>0$ and $\tau_1>0$ such that inequality
$$
    \frac{dV}{d\tau}\Big|_{\eqref{AlmHam}}\leq -\frac{m\sqrt\lambda}{4}\big(R^2+\sigma \Psi^2\big)
$$
holds for any $(R,\Psi,\tau)\in\mathcal B(\rho_1,\tau_1)$.
Similarly, there exists $\rho_2>0$, $\tau_2>0$ such that
\begin{equation}\label{LFE}{\displaystyle
    \frac{1}{4} (R^2+\sigma \Psi^2)\leq V(R,\Psi,\tau)\leq \frac{3}{4} (R^2+\sigma \Psi^2), \quad \forall\, (R,\Psi,\tau)\in \mathcal B(\rho_2,\tau_2).
    }
\end{equation}
Thus
\begin{equation}\label{Est1}{\displaystyle
    \frac{dV}{d\tau}\Big|_{\eqref{AlmHam}}\leq -\frac{m\sqrt\lambda}{3} V
}
\end{equation}
for each triple  $(R,\Psi,\tau)\in\mathcal B(\rho_0,\tau_0)$, where $\rho_0=\min\{\rho_1,\rho_2\}$ and $\tau_0=\max\{\tau_1,\tau_2\}$.
Let $\epsilon$ be an arbitrary positive constant such that $0<\epsilon<\rho_0$; then
\begin{equation}\label{SupInf}
{\displaystyle
    \sup_{\rho\leq \delta_\epsilon, \tau>\tau_0} V(R,\Psi,\tau)
        \leq
     \frac{3\delta^2_\epsilon}{4}
        <
     \frac{\sigma \epsilon^2}{4}
        \leq
     \inf_{\rho=\epsilon, \tau>\tau_0} V(R,\Psi,\tau), \quad \delta_\epsilon=\epsilon\sqrt{\frac{\sigma}{6}}.
}\end{equation}
Therefore any solution $R(\tau)$, $\Psi(\tau)$ of system \eqref{AlmHam} with initial data $[R^2(\tau_0)+\Psi^2(\tau_0)]^{1/2}\leq \delta_\epsilon$ cannot leave $\epsilon$-neighborhood of the equilibrium $(0;0)$ as $\tau>\tau_0$: $[R^2(\tau)+\Psi^2(\tau)]^{1/2}<\epsilon$.
Integrating \eqref{Est1} with respect to $\tau$, we obtain
$$
    0\leq V(R,\Psi,\tau)\leq C \exp(-2 l \tau) \quad \forall (R,\Psi,\tau)\in\mathcal B(\rho_0,\tau_0),
$$
where positive constant $C$ depends on a trajectory of system \eqref{AlmHam}, $l=m\sqrt\lambda/6>0$.
Hence the Lyapunov function tends exponentially to zero along the trajectories of system \eqref{AlmHam}.
If we combine this with \eqref{Exch1} and \eqref{LFE}, we obtain asymptotic estimates for solutions of system \eqref{Main_Res} with initial data from a neighborhood of the solution $R_-(\tau)$, $\Psi_-(\tau)$:
$$
    r(\tau)=R_-(\tau)+\mathcal O(\tau^{1/2}e^{-l\tau}), \quad \psi(\tau)=\Psi_-(\tau)+\mathcal O(e^{-l\tau}), \quad \tau>\tau_0.
$$
This completes the proof.

Actually, the sufficient conditions obtained in Theorem~\ref{Th1} are almost necessary; this can be seen from the following theorem.
\begin{Th}
If $0<\delta<1$ and $f<0$, then the solution $R_-(\tau)$, $\Psi_-(\tau)$ with asymptotics \eqref{As} is unstable.
\end{Th}
\noindent
The {\bf proof} is based on Lyapunov function \eqref{LF}. Since $f<0$, we have
$$
    \frac{dV}{d\tau}\Big|_{\eqref{AlmHam}}\geq -\frac{\delta f}{4}(R^2+\sigma\Psi^2)>0
$$
for any $(R,\Psi,\tau)\in\mathcal B(\rho_0,\tau_0)$. From the Lyapunov theorem (see~\cite{Malkin52}) it follows that the solution $R_-(\tau)$, $\Psi_-(\tau)$ is unstable.

\section{Deterministic perturbations}

In this section we consider the problem of stability of the capture into the parametric autoresonance under persistent deterministic perturbations. Our goal is to identify a class of functions $(\xi,\eta,\zeta)$ such that system \eqref{PSD} has resonance solutions with growing energy as the perturbation parameter $\mu$ is sufficiently small.

Let $\mathcal T_\mu$ be a positive function such that $\mathcal T_\mu\to\infty$ as $\mu\to 0$.
Now we shall give the following definition of stability (see~\cite{Hapaev88}).
\begin{Def}
The solution $R_-(\tau)$, $\Psi_-(\tau)$ of system \eqref{Main_Res} is stable under persistent perturbations $\mathcal{P}$ on an asymptotically large interval $(\tau_0; \tau_0+\mathcal T_\mu)$ if
$\forall\, \epsilon>0$  $\exists \, \delta_\epsilon, \Delta_\epsilon>0$ {\rm :}
$$
    \forall \,\varrho_0, \phi_0: \ \  |\varrho_0-R_-(\tau_0)|+|\psi_0-\Psi_-(\tau_0)|\leq\delta_\epsilon, \ \ \forall \mu<\Delta_\epsilon, \quad \forall (\xi,\eta,\zeta)\in\mathcal P
$$
 the solution $r_\mu(\tau)$, $\psi_\mu(\tau)$ of system \eqref{PSD} with initial data $r_\mu(\tau_0)=\varrho_0$,  $\psi_\mu(\tau_0)=\phi_0$ satisfies the inequality
$$
    \sup_{0<\tau-\tau_0< \mathcal T_\mu}|r_\mu(\tau)-R_-(\tau)| \tau^{-1/2} + |\psi_\mu(\tau)-\Psi_-(\tau)|< \epsilon.
$$
\end{Def}

Note that this definition is different from a classical one~\cite{Malkin52} because of the finite time interval. But such an approach seems to be reasonable, since the considered mathematical model \eqref{Main_Res} is valid only for $0<\tau\ll\varepsilon^{-1}$, where $0<\varepsilon\ll 1$ is the driving amplitude in the nonlinear systems like \eqref{Duffing}. In order to describe the autoresonance phenomenon for $\tau\gg \varepsilon^{-1}$ one should consider other equations~\cite{LK08}.

Consider a class $\mathcal D_{a,b,c}$ of functions $(\xi,\eta,\zeta)$ such that $\forall (\xi,\eta,\zeta)\in\mathcal D_{a,b,c}$
$$
    \sup_{(r,\psi)\in\mathbb R^2,\tau>0} |\xi(r,\psi,\tau)|\tau^{-a}+|\eta(r,\psi,\tau)|\tau^{-b}+|\zeta(r,\psi,\tau)|\tau^{-c}<\infty.
$$
Let $h>0$ be a positive constant; we define a class $\mathcal D_{a,b,c}^h$ as a subset of $\mathcal D_{a,b,c}$ such that for any $(\xi,\eta,\zeta)\in\mathcal D_{a,b,c}^h$ $$\sup_{(r,\psi)\in\mathbb R^2,\tau>0} |\xi(r,\psi,\tau)|\tau^{-a}+|\eta(r,\psi,\tau)|\tau^{-b}+|\zeta(r,\psi,\tau)|\tau^{-c}\leq h.$$
The Cauchy problem for perturbed system \eqref{PSD} with initial data from a neighborhood of the solution $R_-(\tau)$, $\Psi_-(\tau)$ is assumed to have a global solution. This requirement impose the additional restrictions on the class of perturbations (see~\cite{NemStep04}).

\begin{Th}\label{Th3}
If $0<\delta<1$, $f>0$, then $\forall\, h>0$, $a>-1/2$, $b>0$, $c>0$, and $\varkappa\in(0;\varkappa_0)$ the solution $R_-(\tau)$, $\Psi_-(\tau)$ with asymptotics \eqref{As} is stable under the persistent perturbations $(\xi,\eta,\zeta)\in\mathcal D^h_{a,b,c}${\rm :}
$$
    \sup_{(r,\psi)\in\mathbb R^2,\tau>0} |\xi(r,\psi,\tau)|\tau^{-a}+|\eta(r,\psi,\tau)|\tau^{-b}+|\zeta(r,\psi,\tau)|\tau^{-c}\leq h
$$
on an asymptotically large interval $0<\tau< \mathcal O(\mu^{-\varkappa})$, where $\varkappa_0=\vartheta^{-1}$, $\vartheta=\max\{a+1/2,b,c\}$.
\end{Th}
\noindent
{\bf Proof.}
As above, we reduce the considered problem to the analysis of the equilibrium $(0;0)$ in system \eqref{AlmHam}.
By change of variables \eqref{Exch1} perturbed system \eqref{PSD} is reduced to the differential equations
\begin{equation}
\label{AlmHamPert}
\begin{array}{c}
    {\displaystyle
       \frac{1}{\sqrt{\lambda \tau}} \frac{dR}{d\tau}=-\partial_\Psi H +F+\mu G, \quad
       \frac{1}{\sqrt{\lambda \tau}} \frac{d\Psi}{d\tau}=\partial_R H+\mu Q,
    }
\end{array}
\end{equation}
where
\begin{equation}
\begin{array}{l}\nonumber
        {\displaystyle
            G(R,\Psi,\tau)=  (R_-+R\sqrt{\lambda\tau})\sin(\Psi+\Psi_-)\frac{\hat \xi}{\lambda\tau}},
            \quad
        {\displaystyle
            Q(R,\Psi,\tau)= (f \hat \eta\cos(\Psi+\Psi_-)+\hat \zeta)\frac{1}{\sqrt{\lambda\tau}}.
        }
\end{array}\end{equation}
Persistent perturbations of system \eqref{AlmHam} are associated with the functions $G$ and $Q$.
The functions $\hat\xi(R,\Psi,\tau)$, $\hat\eta(R,\Psi,\tau)$, and $\hat\zeta(R,\Psi,\tau)$ are associated with $\xi(r,\psi,\tau)$, $\eta(r,\psi,\tau)$, and $\zeta(r,\psi,\tau)$ through \eqref{Exch1}, e.g.,
$\hat\xi(R,\Psi,\tau)=\xi(R_-(\tau)+ R \sqrt{\lambda\tau}, \Psi_-(\tau)+\Psi, \tau)$.

Let $h>0$, $a>-1/2$, $b>0$, and $c>0$ be arbitrary constants. The derivative of the Lyapunov function \eqref{LF} with respect to $\tau$ along the trajectories of perturbed system \eqref{AlmHamPert} has the form:
\begin{equation}
    \label{TD}{\displaystyle
        \frac{dV}{d\tau}\Big|_{\eqref{AlmHamPert}}=\frac{dV}{d\tau}\Big|_{\eqref{AlmHam}}+\mu \sqrt{\lambda\tau}(G\, \partial_R V +Q\,\partial_\Psi V).
        }
\end{equation}
Note that the first term in the right-hand side of \eqref{TD} satisfies the inequality \eqref{Est1} in the domain $\mathcal B(\rho_0,\tau_0)$, while the derivatives $\partial_R V$, $\partial_\Psi V$ are bounded: $|\partial_R V|+|\partial_\Psi V|\leq \ell$.
From the definition of the class $\mathcal D^h_{a,b,c}$ it follows that
$$
   \sqrt{\tau} (|G|+|Q|)\leq   M_h \tau^{\vartheta}, \quad (R,\Psi,\tau)\in\mathcal B(\rho_0,\tau_0),
$$
where $M_h$ is the positive constant and $\vartheta>0$.
Thus the inequality
$$\frac{dV}{d\tau}\Big|_{\eqref{AlmHamPert}}\leq -\frac{m\sqrt\lambda}{3}\Big[ V  - \mu\, \frac{6 \ell M_h}{m}\, \tau^{\vartheta}\Big] $$
holds for any $(R,\Psi,\tau)\in\mathcal B(\rho_0,\tau_0)$.
For any $\epsilon$ and $\varkappa$ such that $0<\epsilon<\rho_0$ and $0<\varkappa<1/\vartheta$ we define
$$
    \delta_\epsilon=\epsilon\sqrt{\frac{\sigma}{6}}, \quad
    \Delta_\epsilon=\Big[\frac{\sigma \delta_\epsilon^2 m }{24 (2\tau_0)^{\vartheta}\ell M_h }\Big]^{1/z}, \quad z=1-\vartheta\varkappa > 0.
$$
Then the derivative of the function $V$ is negative:
\begin{equation}\nonumber {\displaystyle
    \frac{dV}{d\tau}\Big|_{\eqref{AlmHamPert}}\leq -\frac {m\sqrt\lambda}{3} \Big[V-\frac{\sigma \delta_\epsilon^2}{4}\Big]\leq 0
}\end{equation}
if $\mu<\Delta_\epsilon$, $\delta_\epsilon<\rho<\rho_0$ and $0<\tau-\tau_0\leq \tau_0 \mu^{-\varkappa}$.
Let us remember that the Lyapunov function satisfies inequalities \eqref{SupInf}. Hence any solution $R_\mu(\tau)$, $\Psi_\mu(\tau)$ of perturbed system \eqref{AlmHamPert} such that $[R_\mu^2(\tau_0)+\Psi_\mu^2(\tau_0)]^{1/2}\leq \delta_\epsilon$ cannot leave the $\epsilon$-neighborhood of the equilibrium $(0;0)$ as $0<\tau-\tau_0\leq \tau_0 \mu^{-\varkappa}$. Taking into account \eqref{Exch1}, we obtain
$$
    |r_\mu(\tau)-R_-(\tau)|(\lambda \tau)^{-1/2}+|\psi_\mu(\tau)-\Psi_-(\tau)|< \epsilon,\quad 0<\tau-\tau_0\leq \mathcal O(\mu^{-\varkappa} ).
$$
From~\cite{NemStep04} it follows that the solution $R_-(\tau)$, $\Psi_-(\tau)$ is stable on the finite interval $(0;\tau_0]$.
Therefore for any $h>0$, $a>-1/2$, $b>0$, $c>0$, and $\varkappa\in(0; \varkappa_0)$ the solution $R_-(\tau)$, $\Psi_-(\tau)$ of system \eqref{Main_Res} is stable under the persistent perturbations $(\xi,\eta,\zeta)\in\mathcal D^h_{a,b,c}$ on an asymptotically large interval
$0<\tau \leq \mathcal O(\mu^{-\varkappa})$.

\begin{Th}
If\, $0<\delta<1$, $f>0$, then $\forall\, h>0$, $a\leq-1/2$, $b\leq0$, and $c\leq0$ the solution $R_-(\tau)$, $\Psi_-(\tau)$ with asymptotics \eqref{As} is stable under persistent perturbations $(\xi,\eta,\zeta)\in\mathcal {D}^h_{a,b,c}${\rm :}
$$
    \sup_{(r,\psi)\in\mathbb R^2,\tau>0} |\xi(r,\psi,\tau)|\tau^{-a}+|\eta(r,\psi,\tau)|\tau^{-b}+|\zeta(r,\psi,\tau)|\tau^{-c}\leq h
$$
on the infinite interval $\tau>0$.
\end{Th}
\noindent
The {\bf proof} is follows from Malkin's theorem~\cite{Malkin52}.

\section{Random perturbations}
We consider the perturbed system:
\begin{equation}\label{PSR}
    \begin{array}{c}
        {\displaystyle \frac{dr}{d \tau}=(1+\xi)r\sin\Phi-\delta r, \ \  \frac{d\psi}{d \tau}=r-\lambda \tau+ \zeta+(f+ \eta)\cos\psi,}
    \end{array}
\end{equation}
where $\xi(r,\psi,\tau;\omega,\mu)$, $\eta(r,\psi,\tau;\omega,\mu)$, $\zeta(r,\psi,\tau;\omega,\mu)$ are one-dimensional random processes defined on a probability space $(\Omega, \mathcal F, {\bf P})$, $(r,\psi)\in \mathbb R^2$, $\tau>0$, $\mu>0$.
Our goal is to find a class of perturbations $(\xi,\eta,\zeta)$ such that the stability of autoresonance is preserved in perturbed system \eqref{PSR}.

Let $\mathcal T_\mu$ be a positive function such that $\mathcal T_\mu\to\infty$ as $\mu\to 0$.  Now we give the following definition of stability under random perturbations.

\begin{Def}
The solution $R_-(\tau)$, $\Psi_-(\tau)$ of system \eqref{Main_Res} is stable under random perturbations $\mathcal{P}$ on an asymptotically large interval $(\tau_0; \tau_0+\mathcal T_\mu)$ if  $\forall\, \epsilon, \upsilon>0$ $\exists \, \delta_\epsilon,\Delta>0$ {\rm :}
$$ \forall \,\varrho_0, \phi_0: \ \  |\varrho_0-R_-(\tau_0)|+|\psi_0-\Psi_-(\tau_0)|\leq\delta_\epsilon, \ \ \forall \mu<\Delta, \ \ \forall (\xi,\eta,\zeta)\in\mathcal P$$
 the solution $r_\mu(\tau;\omega)$, $\psi_\mu(\tau;\omega)$ of system \eqref{PSR} with initial data $r_\mu(\tau_0;\omega)=\varrho_0$,  $\psi_\mu(\tau_0;\omega)=\phi_0$  satisfies the inequality
$${\bf P}\Big(\sup_{0<\tau-\tau_0<\mathcal T_\mu}|r_\mu(\tau;\omega)-R_-(\tau)| \tau^{-1/2} + |\psi_\mu(\tau;\omega)-\Psi_-(\tau)|>\epsilon\Big)< \upsilon.$$
\end{Def}
Note that this definition is usually used for strong stability (see \cite[p.~152]{Hasm69}, \cite[Chap.~9]{Hapaev88}, and ~\cite[p.~400]{Schuss10}).

Consider a class $\mathcal {R}_{a,b,c}$ of random functions $\xi,\eta,\zeta$. Assume that for all $(\xi,\eta,\zeta)\in \mathcal {R}_{a,b,c}$ there exists at least one random function $S(\tau;\omega,\mu)$ such that
\begin{equation}\nonumber
    {\displaystyle \exists\, \nu(\omega) >0 : \quad {\bf M}_\tau S\stackrel{def}{=} \int\limits_{\tau}^{\tau+1} S(t;\omega,\mu)\, dt\leq \mu\,\nu(\omega) \quad \forall \, \tau>0, \quad \omega\in\Omega, \quad \mu>0,}
\end{equation}
and
\begin{equation}\nonumber
{\displaystyle{\bf E} \, \nu\stackrel{def}{=}\int\limits_{\Omega} \nu(\omega) {\bf P}\, (d\omega)<\infty.}
\end{equation}
It is assumed that for all $(\xi,\eta,\zeta)\in\mathcal R_{a,b,c}$
$$\sup_{(r,\psi)\in\mathbb R^2} |\xi(r,\psi,\tau;\omega,\mu)|\tau^{-a}+|\eta(r,\psi,\tau;\omega,\mu)|\tau^{-b}+|\zeta(r,\psi,\tau;\omega,\mu)|\tau^{-c}<S(\tau;\omega)$$
for all $\tau>0$, $\omega\in\Omega$, and $\mu>0$.

Let $h>0$ be a positive constant; we define a class $\mathcal {R}_{a,b,c}^h$ as a subset of $\mathcal {R}_{a,b,c}$ such that ${\bf E}\,\nu\leq h$.

\begin{figure}[H]
\centering
\includegraphics[width=0.4\textwidth]{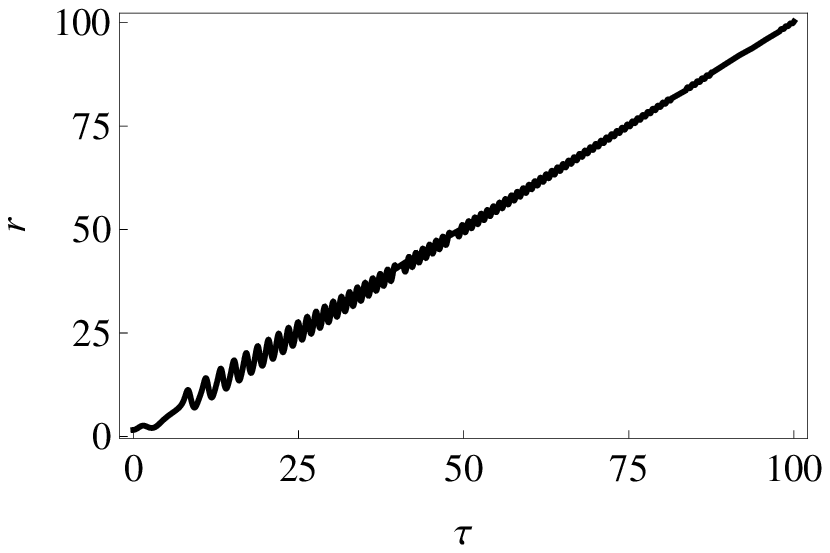} \quad \includegraphics[width=0.4\textwidth]{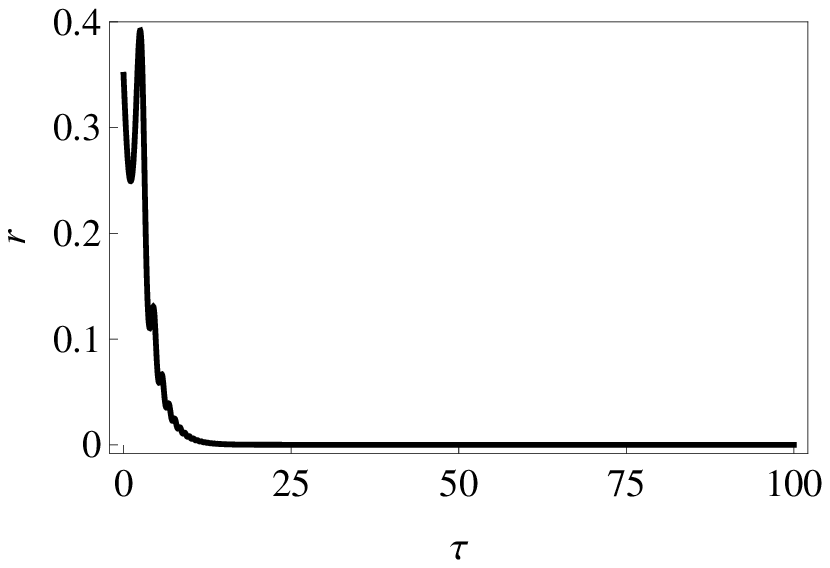}
\caption{ System \eqref{Main_Res} has solutions with bounded and unlimited amplitude. Two curves correspond to different initial data: (left) $r(0) = 1.59$, $\psi(0) = 0.59$, (right) $r(0) = 0.35$, $\psi(0) = 3.09$. Parameter values are  $f = 0.2$, $\delta=0.5$, and $\lambda = 1$.}
\label{TwoP2}
\end{figure}

Notice that the stability under random perturbations with bounded expectation ($\sup_t {\bf E}S <\infty$) was investigated in~\cite[p.~26]{Hasm69} provided the unperturbed system is dissipative in the sense of~\cite[p.~8]{Hasm69}. However, the considered equations do not have this property because system \eqref{Main_Res} has solutions of two types: with bounded and unlimited amplitudes, see Figure \ref{TwoP2}. Thus the results of~\cite{Hasm69} are not applicable. Stability of nondissipative systems is discussed in~\cite{LK12,OS14}. We extend these results to the analysis of stability on an asymptotically large interval.

\begin{Th}\label{Th5}
If $0<\delta<1$, $f>0$, then $\forall\, h>0$, $a>-1/2$, $b>0$, $c>0$, and $\varkappa\in(0;\varkappa_0)$ the solution $R_-(\tau)$, $\Psi_-(\tau)$ with asymptotics \eqref{As} is stable under random perturbations $(\xi,\eta,\zeta)\in\mathcal {R}^h_{a,b,c}${\rm :}
$$\begin{array}{c}
  {\displaystyle  \sup_{(r,\psi)\in\mathbb R^2} |\xi|\tau^{-a}+|\eta|\tau^{-b}+|\zeta|\tau^{-c}\leq S(\tau;\omega,\mu) \quad \forall\, \tau>0, \ \ \omega\in\Omega,  \ \ \mu>0 }  \\
  {\displaystyle  \sup_{\tau>0}{\bf M}_\tau S\leq \mu\,\nu(\omega), \quad  {\bf E}\nu \leq h }
\end{array}
$$
on an asymptotically large interval $0<\tau< \mathcal O(\mu^{-\varkappa})$, where $\varkappa_0=\vartheta^{-1}$, $\vartheta=\max\{a+1/2,b,c\}$.
\end{Th}
\noindent
{\bf Proof.}
We reduce the problem to the analysis of the equilibrium $(0;0)$ in system \eqref{AlmHam}.
Change of variables \eqref{Exch1} leads to the perturbed system
\begin{equation}
\label{AHPR}
\begin{array}{c}
    {\displaystyle
       \frac{1}{\sqrt{\lambda \tau}} \frac{dR}{d\tau}=-\partial_\Psi H +F+G, \quad
       \frac{1}{\sqrt{\lambda \tau}} \frac{d\Psi}{d\tau}=\partial_R H+ Q,
    }
\end{array}
\end{equation}
where random functions $G(R,\Psi,\tau;\omega,\mu)$, $Q(R,\Psi,\tau;\omega,\mu)$ are defined as follows:
$$
\begin{array}{c}
        {\displaystyle
            G=  (R_-+R\sqrt{\lambda\tau})\sin(\Psi+\Psi_-)\frac{ \xi}{\lambda\tau}},
            \quad
        {\displaystyle
            Q= (f \eta\cos(\Psi+\Psi_-)+ \zeta)\frac{1}{\sqrt{\lambda\tau}}.
        }
\end{array}
$$

Let $\epsilon>0$, $\upsilon>0$, $h>0$, $a>-1/2$, $b>0$, and $c>0$ be arbitrary constants.
We construct a Lyapunov function for system \eqref{AHPR} on the basis of the Lyapunov function $V(R,\Psi,\tau)$ for the unperturbed system (see \cite{Krasov63}):
$$U(R,\Psi,\tau;\omega,\mu)=V(R,\Psi,\tau) \exp \Phi(\tau;\omega,\mu),$$
where a smooth function $\Phi(\tau;\omega,\mu)$ is defined bellow.
The derivative of $U(R,\Psi,\tau;\omega,\mu)$ with respect to $\tau$ along the trajectories of perturbed system \eqref{AHPR} has the form:
\begin{equation}\nonumber
        {\displaystyle
        \frac{dU}{d\tau}\Big|_{\eqref{AHPR}}=
        \partial_\tau\Phi\, U +\Big(\frac{dV}{d\tau}\Big|_{\eqref{AlmHam}}+
        \sqrt{\lambda\tau}(G\, \partial_R V +Q\,\partial_\Psi V)\Big)\exp\Phi.
        }
\end{equation}
From the definition of the class $\mathcal R^h_{a,b,c}$ it follows that
$$
    (|G|+|Q|)\tau^{1/2-\vartheta}\leq   q\cdot S(\tau;\omega,\mu), \quad (R,\Psi,\tau)\in\mathcal B(\rho_0,\tau_0),
$$
where $q$ is the positive constant and $\vartheta>0$.
The partial derivatives $\partial_R V$, $\partial_\Psi V$ satisfy the inequality:
$|\partial_R V|+|\partial_\Psi V|\leq \rho\, \ell/\rho_0$ in the domain $\mathcal B (\rho_0,\tau_0)$.
Hence in view of \eqref{LFE} and \eqref{Est1} the derivative of $U(R,\Psi,\tau;\omega,\mu)$ satisfy the estimate:
$$
    \frac{dU}{d\tau}\Big|_{\eqref{AHPR}}\leq \partial_\tau \Phi\, U  -\frac{m\sqrt\lambda}{3}\Big(1-\mu\tau^{\vartheta} \frac{24 \ell q }{m \sigma  \delta_\epsilon \rho_0} \frac{S}{\mu}\Big) U
$$
in the annular domain $\delta_\epsilon<\rho<\epsilon$, $\tau>\tau_0$.
For any $\varkappa$ such that $0<\varkappa<1/\vartheta$ we define a random variable
$$
    \Delta_\omega=\Big[\frac{\Delta_0}{\nu(\omega)}\Big]^z,\quad \Delta_0=\frac{m \sigma \delta_\epsilon \rho_0 }{24\, (2\tau_0)^{\vartheta}\, \ell\, q},
    \quad z=\frac{1}{1-\vartheta\varkappa} > 0.
$$
Then for any $\mu\leq \Delta_\omega$ the inequality
$$
    \frac{dU}{d\tau}\Big|_{\eqref{AHPR}}\leq \partial_\tau \Phi\, U - \frac{m\sqrt\lambda}{3\,\nu}\Big(\nu-\frac{S}{\mu} \Big) U
$$
holds in the domain $\delta_\epsilon<\rho<\epsilon$, $0<\tau-\tau_0<\tau_0 \mu^{-\varkappa}$.

For a fixed $\omega\in\Omega$ we consider the integral
$$I(k;\omega,\mu)=\int\limits_{k}^{k+1}\nu(\omega)-\frac{S(t;\omega,\mu)}{\mu}\, dt, \quad k=0,1,2,\dots.$$
From the definition of the class $\mathcal R_{a,b,c}^h$  it follows that $I(k;\omega,\mu)\geq 0$ for any $k\geq 0$ and $\omega\in\Omega$.
Define an auxiliary random function $\theta(\tau;\omega,\mu)$ such that
\begin{equation}\label{theta}
    {\displaystyle
        \int\limits_{k}^{k+1}\theta(t;\omega,\mu)dt=I(k;\omega,\mu) \quad \forall \, k\geq 0.
    }
\end{equation}
Since right-hand side of \eqref{theta} is not negative, it follows that there exists a non-negative function $\theta(\tau;\omega,\mu)\geq 0$.
Without loss of generality, we can assume that $\theta(\tau;\omega,\mu)$ is a continuous function such that $\theta(k;\omega,\mu)=0$ for any $k=0,1,2,\dots$.
Let us define $\Phi(\tau;\omega)$ as follows
$$
    \Phi(\tau;\omega,\mu)\equiv\frac{m \sqrt{\lambda}}{3} \int\limits_{0}^{\tau} \nu(\omega)-\frac{S(t;\omega,\mu)}{\mu}-\theta(t;\omega,\mu)\, dt.
$$
Then the derivative of $U$ satisfies the inequality:
$$\frac{dU}{d\tau}\Big|_{\eqref{AHPR}}\leq - \frac{m\sqrt\lambda}{3\,\nu }\, \theta \, U\leq 0$$
in the domain $\delta_\epsilon<\rho<\epsilon$, $0<\tau-\tau_0<\tau_0 \mu^{-\varkappa}$.
Taking into account properties of the functions $\theta(\tau; \omega,\mu)$ and $S(\tau; \omega,\mu)$, we obtain $|\Phi(\tau; \omega,\mu)|\leq \Phi_0$, $\Phi_0=4m\sqrt\lambda/3$.
Thus for any $\epsilon>0$, $\omega\in\Omega$, and $\mu>0$ we have
\begin{equation}\nonumber
{\displaystyle
    \sup_{\rho\leq \delta_\epsilon, \tau>\tau_0} U(R,\Psi,\tau;\omega,\mu)
        \leq
     \frac{3\delta^2}{4} \exp \Phi_0
        <
     \frac{\sigma \epsilon^2}{4} \exp(-\Phi_0)
        \leq
     \inf_{\rho=\epsilon, \tau>\tau_0} U(R,\Psi,\tau;\omega,\mu),
}\end{equation}
where $\delta_\epsilon=\epsilon \, \exp(-\Phi_0) \sqrt{\sigma/6}$.
Consider perturbations $(\xi,\eta,\zeta)\in \mathcal R_{a,b,c}^h$ such that $\nu(\omega)\leq h/\upsilon$ uniformly for all $\omega\in\Omega$.
Then the parameter $\mu$ can be bounded away from zero $0<\mu<\Delta=(\upsilon\Delta_0/h)^z\leq \Delta_\omega$.
Therefore any solution $R(\tau;\omega)$, $\Psi(\tau;\omega)$ starting from the neighborhood of equilibrium
$[R_\mu^2(\tau_0;\omega)+\Psi_\mu^2(\tau_0;\omega)]^{1/2}\leq \delta_\epsilon$ remains inside the ball
$[R_\mu^2(\tau;\omega)+\Psi_\mu^2(\tau;\omega)]^{1/2}<\epsilon$ for $0<\tau-\tau_0< \mathcal O(\mu^{-\varkappa})$.

For all other perturbations such that $\nu(\omega)> h/\upsilon$, it follows from the Chebyshev inequality that
$$
    {\bf P}(\nu(\omega)> h/\upsilon)<\frac{{\bf E}\,\nu}{h/\upsilon}\leq\upsilon.
$$
In this case, solutions of perturbed system \eqref{AlmHamPert} can leave any $\epsilon$-neighborhood of the equilibrium $(0;0)$.
However, the probability of such events is small:
$$
    {\bf P}(\sup_{0<\tau-\tau_0< \mathcal T_\mu} [R_\mu^2(\tau;\omega)+\Psi_\mu^2(\tau;\omega)]^{1/2}>\epsilon )<\upsilon, \quad \mathcal T_\mu=\tau_0\mu^{-\varkappa}.
$$

The change-of-variables formula \eqref{Exch1} implies the stability of the solution $R_-(\tau)$, $\Psi_-(\tau)$
of system \eqref{Main_Res} under random perturbations on the asymptotically large interval $0<\tau-\tau_0 < \mathcal O(\mu^{-\varkappa})$.
Stability on the interval $\tau\in(0;\tau_0]$ follows from the continuity of solutions with respect to parameters, see~\cite{Hasm69}.
This concludes the proof.

\begin{Th}
If\, $0<\delta<1$, $f>0$, then $\forall\, h>0$, $a\leq-1/2$, $b\leq0$, and $c\leq0$ the solution $R_-(\tau)$, $\Psi_-(\tau)$ with asymptotics \eqref{As} is stable under random perturbations $(\xi,\eta,\zeta)\in\mathcal {R}^h_{a,b,c}${\rm :}
$$
    \begin{array}{c}
        {\displaystyle  \sup_{(r,\psi)\in\mathbb R^2} |\xi|\tau^{-a}+|\eta|\tau^{-b}+|\zeta|\tau^{-c}\leq S(\tau;\omega,\mu) \quad \forall\, \tau>0, \ \ \omega\in\Omega, \ \ \mu>0, }  \\
        {\displaystyle  \sup_{\tau>0}{\bf M}_\tau S \leq \mu \nu(\omega), \quad  {\bf E}\, \nu\leq h }
    \end{array}
$$
on the infinite interval $\tau>0$.
\end{Th}
\noindent
The {\bf proof} follows from the results of \cite{LK12} and \cite{OS14}.

\section{Examples}

{\bf 1.} To illustrate Theorem \ref{Th3}, let us consider the following example.
Let $0<\delta<1$, $f>0$, $\xi(\tau)\equiv 1$, $\eta(\tau)\equiv 1$, and $\zeta(\tau)=\tau$. Then perturbed system \eqref{PSD} takes the form:
$$
        \frac{dr}{d\tau} = (1+\mu) r \sin\psi - \delta r, \quad \frac{d\psi}{d\tau} = r - \lambda \tau + \mu\tau + (f + \mu) \cos\psi,  \quad 0<\mu\ll 1.
$$
Note that one can easily construct the asymptotic expansion at infinity $\tau\to\infty$ for a particular resonant solution of the perturbed system as follows
$$
    r_\mu(\tau)=(\lambda-\mu)\tau+\mathcal O(1), \quad \psi_\mu(\tau)=\pi-\arcsin\Big(\frac{\delta}{1+\mu}\Big)+\mathcal O(\tau^{-1}).
$$
Therefor we have
$$
    |R_-(\tau)-r_\mu(\tau)|\tau^{-1/2}=\mu \tau^{1/2}[1+\mathcal O(\tau^{-1})], \quad |\Psi_-(\tau)-\psi_\mu(\tau)|=\mathcal O(\mu)
$$
as $\tau\to\infty$ and $\mu \to 0$. From the first estimate it follows that the solution $R_-(\tau)$, $\Psi_-(\tau)$ is unstable for $\tau>\mu^{-2}$. However, since $\xi,\eta,\zeta$ belong to $\mathcal D_{0,0,1}^1$, we see that from Theorem \ref{Th3} it follows that the solution $R_-(\tau)$, $\Psi_-(\tau)$ is stable under the persistent perturbations at the interval $0<\tau< \mathcal O(\mu^{-\varkappa})$, where $0<\varkappa<1$.

{\bf 2.} Let us consider random perturbations of the prime parametric resonance equations. First define the random process
$$
    J_N(\tau;\omega,\mu)\equiv \sum\limits_{n=1}^{N}j_n(\omega) \chi{(n\leq \tau\leq n+\mu )}, \quad 0<\mu\ll 1, \quad \tau\geq 0, \quad \omega\in\Omega, \quad N\in\mathbb N,
$$
where $\chi$ is a characteristic function, $\{j_n(\omega)\}_{n=1}^{N}$ are random variables such that ${\bf E}\,| j_n |<\infty$.
Since ${\bf M}_\tau | J_N | \leq \mu [ |j_n|+ |j_{n+1}|]$ for $\tau \in [n; n+1)$,  we see that the random functions
$$
    \xi(\tau;\omega,\mu)\equiv J_N(\tau;\omega,\mu), \quad
    \eta(\tau;\omega,\mu)\equiv J_N(\tau;\omega,\mu),\quad
    \zeta(\tau;\omega,\mu)\equiv \tau J_N(\tau;\omega,\mu),
$$
belong to the class $\mathcal R_{0,0,1}$.
From theorem~\ref{Th5} it follows that the solution $R_-(\tau)$, $\Psi_-(\tau)$ is stable as $\tau\ll \mu^{-1}$.

{\bf 3.} Another example of perturbations is described by the function with a random jump. Let
$$
    \xi(\tau;\omega,\mu)\equiv J(\tau;\omega,\mu), \quad
    \eta(\tau;\omega,\mu)\equiv \tau J(\tau;\omega,\mu),\quad
    \zeta(\tau;\omega,\mu)\equiv\tau J(\tau;\omega,\mu),
$$
where $J(\tau;\omega,\mu)\equiv j(\omega) \chi{(\omega\leq \tau\leq\omega+\mu)}$, $0<\mu\ll 1$, $\tau\geq 0$, $\omega\in\Omega=(0,\infty)$, and $j(\omega)\not\equiv0$ is a random variable with the bounded expectation ${\bf E} | j |<\infty$.
Since ${\bf M}_\tau |J(\tau;\omega,\mu)|\leq \mu |j(\omega)|$ for $\tau\geq 0$, it follows that $(\xi,\eta,\zeta)\in \mathcal R_{0,1,1}$ and the solution $R_-(\tau)$, $\Psi_-(\tau)$ is stable as $\tau\ll \mu^{-1}$.

Finally note that from Theorem~\ref{Th3} and \eqref{PF} it follows that stability of the capture into autoresonance in perturbed system \eqref{PDuf} is preserved on the interval $0<\varepsilon t\ll \mu^{-1}$ if
$$
    0<\beta<\frac{\varepsilon}{2}, \quad |a(t;\varepsilon)|\leq \mu\sqrt{\varepsilon t}, \quad |\varphi(t;\varepsilon)|\leq \mu \varepsilon^2 t^2, \quad t\geq 0.
$$

\section{Conclusion}

Stability of the capture into the parametric autoresonance in dissipative systems under persistent perturbations on a long time interval was proved. The length of the interval depends on both the perturbation parameter $\mu$ and the class $\mathcal P$ of admissible perturbations.
Stability of autoresonance under white noise perturbations was not considered. This will be discussed in a further paper.

\section*{Acknowledgments}

This work was supported by Russian Science Foundation, project no. 14-11-00078.

\end{document}